\documentclass[aps, reprint, superscriptaddress, nofootinbib]{revtex4-1}
 \pdfoutput=1
\usepackage{hyperref} 
\usepackage{amsmath, latexsym, amssymb, graphicx, color, slashed}
\definecolor{nicered}{rgb}{0.7,0.1,0.1}
\definecolor{nicegreen}{rgb}{0.1,0.5,0.1}
\hypersetup{colorlinks, citecolor=nicegreen,linkcolor=nicered}

\begin{document}

 \newcommand{\Sec}[1]{ \medskip \noindent {\sl \bfseries #1}}
\newcommand{\subsec}[1]{ \medskip \noindent {\sl \bfseries #1}}
\newcommand{\Par}[1]{ \medskip \noindent {\em #1}}

\title{Right Handed Quark Mixing in Left-Right Symmetric Theory}

\author{Goran Senjanovi\'{c}}
\affiliation{Gran Sasso Science Institute, Viale Crispi, L'Aquila, Italy}
\affiliation{International Centre for Theoretical Physics, Trieste, Italy }
\author{Vladimir Tello}
\affiliation{Gran Sasso Science Institute, Viale Crispi, L'Aquila, Italy}
\date{\today}

\begin{abstract}
 
  We give exact formulas for the right-handed analog of the CKM matrix in the minimal Left-Right symmetric theory, for the case when the Left-Right symmetry is generalized Parity as in the original version of the theory. We derive its explicit form and give a physical reason for the known and surprising fact that the right-handed mixing angles are close in value to the CKM ones, in spite of the Left-Right symmetry being badly broken in nature. We exemplify our results on the production of the 
 right-handed charged gauge boson and the computation of $K_L - K_S$ mass difference. 
\end{abstract}

\maketitle

%
%
\Sec{I.  Introduction.} 
        The Left-Right (LR) symmetric theory~\cite{lrmodel} prophetically implied  non-vanishing  neutrino mass whose smallness, through the seesaw mechanism~\cite{Minkowski:1977sc,Mohapatra:1979ia,seesawso10&fam}, is related~\cite{Minkowski:1977sc,Mohapatra:1979ia} to  parity violation at low energies. The theory leads to neutrinoless double beta decay~\cite{Racah:1937qq} through both left handed (LH) and heavy right handed (RH) neutrinos~\cite{Mohapatra:1979ia}. One can in principle observe lepton number violation at hadronic colliders and probe directly the Majorana nature of heavy neutrinos through the so-calleed KS process~\cite{Keung:1983uu}. Moreover, the knowledge of neutrino masses allows one to predict the Dirac Yukawa couplings~\cite{Nemevsek:2012iq} and the associated decays of RH neutrinos.
           
        The small $K_L - K_S$ mass difference implies a lower limit~\cite{soni} on the LR scale in the minimal model around 3 TeV~\cite{Bertolini:2014sua}, and the LHC has come close to it for some channels~\cite{Khachatryan:2014dka}.  
        This limit could go up to 20 TeV~\cite{Xu:2009nt}, but that depends on the UV completion of the theory.

        Detailed studies~\cite{Ferrari:2000sp} support the feasibility of the KS process at the LHC connected to neutrinoless double beta decay and lepton flavor violation~\cite{Tello:2010am}. Recently CMS reported a 2.8 sigma excess~\cite{Khachatryan:2014dka} in the KS process that could be a manifestation of the LR symmetry~\cite{Deppisch:2014qpa}. It would require, however, the RH gauge coupling to be appreciably smaller that the LH one, not discussed here.
                                             
         In the limit of unbroken LR symmetry,  left and right mixings are equal.
        The situation after the LR symmetry breaking depend on its nature,   
 which can be either generalized charge conjugation $\mathcal{C}$ or generalized parity $\mathcal{P}$.
In the case of $\mathcal{C}$,  
 quark mass matrices are symmetric and  the mixing angles remain the same, the only difference lying in  phases. 
            
       
            
             In the case of $\mathcal{P}$,  quark mass matrices are neither symmetric nor hermitian in general, yet 
             it turns out that the
             left and right mixing angles are 
              close to each other, as shown first  numerically 
             in a portion of the parameter space in \cite{Kiers:2002cz}.
              An analytical study  in the same approximation was made in \cite{md}, and \cite{Maiezza:2010ic} established this result over the entire parameter space by combining analytical and numerical computations. 
          
          In this Letter we shed new light on this old issue by finding the explicit form of the right-handed quark mixing matrix,    essential for  high precision phenomenological studies of the  theory. 
          In the process we offer a physical explanation behind the 
          approximate equality of  left and right mixing angles.
           We apply our findings 
           to the production strength of the heavy RH charged gauge boson and 
         to   the computation of the  $K_L - K_S$ mass difference.

\Sec{II. The right handed quark mixing matrix.}
   The LR symmetric theory studied here is based on the $SU(2)_L \times SU(2)_R \times U(1)_{B-L}$ gauge group augmented with a generalized parity $\mathcal{P}: q_L \leftrightarrow q_R$, where $q_{L,R}$ are quark doublets under $SU(2)_L$ and $SU(2)_R$ gauge groups respectively.
   
   The quark Yukawa couplings in the minimal theory take the following form 
\begin{equation}\label{eq:quarks&phi}
L_Y=  \overline{q_{L}}\,\big(Y_{1} \Phi - Y_{2}\,  \sigma_2 \Phi^* \sigma_2)\, q_R+\text{h.c.}
\end{equation}
where $\Phi$ is a Higgs scalar bi-doublet with the non-vanishing vevs%
\begin{equation}
\langle\Phi\rangle=v\, \text{diag} (\cos\beta,-\sin\beta e^{-ia})
\end{equation}
and $ \beta<\pi/4$, $0 < a < 2 \pi$. 

The underlying LR symmetry in the form of generalized parity $\mathcal{P}$ implies hermitian Yukawa matrices, which in turn lead to the following relations between the up and down quark mass matrices
\begin{align}\label{relationsMuMd-1}
M_u-M_u^{\dagger}&=-i s_at_{2\beta} (e^{-ia}t_{\beta}M_u+M_d)\\ \label{relationsMuMd-2}
M_d-M_d^{\dagger}&=is_at_{2\beta}(M_u+e^{i a}t_{\beta}M_d)
\end{align}
where $s_a \equiv \sin a$, $t_\beta \equiv \tan \beta$, $t_{2\beta} \equiv \tan 2\beta$.
From \eqref{relationsMuMd-2}, it is easy to see a rough upper limit $s_a t_{2\beta} \lesssim 2 m_b/m_t$, found before in~\cite{Maiezza:2010ic}.  Clearly, these relations and their consequences are valid above the scale of the LR symmetry breaking and one may need to take into account their scale dependence in order to apply them at low energies.

The mixing matrices arise from  a product of matrices that diagonalize the quark mass matrices
 \begin{equation}
  M_u=U_L m_u U_R^{\dagger},\quad M_d=D_L m_d D_R^{\dagger}
\end{equation}
where $m_q$ are diagonal matrices of positive quark masses. This gives the left-handed CKM matrix $V_L$ and its right-handed analog $V_R$
\begin{equation}
  V_L=U_L^{\dagger}D_L,\,\,\,  V_R=U_R^{\dagger}D_R
 \end{equation}
 It will be  useful to introduce unitary matrices $U_d$ and $U_u$  which become unity (up to sign matrices discussed below) when the corresponding  mass matrices are hermitian 
   \begin{equation}
    U_u=   U_L^{\dagger}U_R,   \quad  U_d=   D_L^{\dagger}D_R  
 \end{equation}
Then from \eqref{relationsMuMd-1} and \eqref{relationsMuMd-2} one finds 
 \begin{align}\label{eq:Uu}
&  U_u=\frac{1}{m_u}\sqrt{m_u^2+is_ a t_{2\beta}   \left(t_{\beta}e^{-i a}m_u^2+m_uV_L m_d V_R^{\dagger}\right) }  
\\
 \label{eq:Ud}
& U_d=\frac{1}{m_d}\sqrt{m_d^2- i s_ a t_{2\beta}  \left(t_{\beta}e^{ia}m_d^2+m_dV_L^{\dagger}m_uV_R\right) }
\end{align}
Additionally, one has a relation which arises from the definition of the mixing matrices
\begin{equation} \label{eq:VR}
V_L U_d=U_u  V_R
\end{equation} 
 Together with \eqref{eq:Uu} and \eqref{eq:Ud} it allows for the determination of $V_R$ in term of $V_L,m_u,m_d,a$ and  $\beta$. 
This tough computational task is simplified, for we can expand in small $s_a t_{2\beta}$. We spare the reader the  details that go into the following leading term expression 
\begin{widetext}
\begin{equation}\label{eq:master}
(V_R)_{ij}= (V_L)_{ij} - i   s_a t_{2\beta}\bigg[  t_{\beta}  (V_L) _{ij}
+ \frac{  (V_Lm_d V_L^{\dagger} )_{ik}  (V_L )_{kj} }{m_{u_i}+m_{u_k}}  +  \frac{(V_L)_{ik} ( V_L^{\dagger}m_uV_L)_{kj} }{m_{d_k}+m_{d_j}} \bigg]  
+O(s_a^2t^2_{2\beta}) 
\end{equation}
\end{widetext}

There are $2^{(2n-1)}$ independent solutions for $n$ generations, due to the square root nature of \eqref{eq:Uu} and \eqref{eq:Ud}. The rest is found through $V_L\rightarrow S_uV_L S_d$ and $m_{q_i}\rightarrow s_{q_i} m_{q_i}$, where $S_u=\text{diag}(s_{u_i})$, $S_d=\text{diag}(s_{d_i})$ and $s_{q_i}$ are $\pm$ signs.

Equation \eqref{eq:master} eloquently expresses the RH mixing matrix $V_R$ as an expansion in the small parameter $s_a t_{2\beta}$, which must satisfy
\begin{equation}
|s_{a}t_{2\beta}|\lesssim \text{min}  \bigg|\dfrac{ m_{d_i}+m_{d_j}}{(V_L^{\dagger}m_{u}V_L)_{ij}} \bigg| \simeq  2 \frac{m_b}{m_t}
\end{equation}
This rigorously confirms what we estimated in the beginning by just taking the third generation, the reason being that the third generation CKM mixing angles are tiny.

Our task is basically completed; it suffices to keep in mind that when $s_a t_{2\beta}$ is close to $ 2 m_b/m_t$, one may have to include higher order terms. This is straightforward and we leave it for future work. What is essential is that we have complete control of the situation through a well-defined expansion procedure. 

Already at this level it can be shown  that the difference between the right and left mixing angles is small, as found numerically in the past. In order to do so, we use the  following parametrisation for a general $ 3 \times 3$ unitary matrix  
\begin{equation}    
V_R\! \equiv\! \text{diag}(e^{i\omega_1}\!,e^{i\omega_2}\!,e^{i\omega_3}\!)V(\theta_{ij}^R,\delta_R) \text{diag}(e^{i\omega_4}\!,e^{i\omega_5}\!,1)
\end{equation}
where  $V(\theta^{ij}_L,\delta_L) \equiv V_L$   is the standard form used by the PDG of the left-handed CKM matrix. Besides the RH analog $\delta_R$  of the KM phase $\delta_L$, $V_R$ contains five extra (external) phases that cannot be rotated away since we used all the phase freedom in defining the usual CKM matrix in the left sector. 

A straightforward computation from \eqref{eq:master} gives the leading terms for the differences between mixing angles and the KM phases
    \begin{align}\label{eq:difference12}
   \theta^{12}_R - \theta^{12}_L &\simeq  - s_a t_{2\beta} \frac {m_t}{m_s+m_d} \sin\theta^{23}_L \sin\theta^{13}_L   \sin \delta_L
\\[3pt]
   \label{eq:difference23}
   \begin{split}
\theta^{23}_R - \theta^{23}_L &\simeq   s_a t_{2\beta}  \left(\frac{m_t}{m_b+m_s}-\frac{m_t}{m_b+m_d}\right) \\ 
&\times \sin\theta^{12}_L \sin\theta^{13}_L  \sin\delta_L
\end{split}
\\[3pt]
    \label{eq:difference13}
    \begin{split}
     \theta^{13}_R - \theta^{13}_L &\simeq  s_a t_{2\beta}\left(\frac{m_t}{m_b+m_d}-\frac{m_t}{m_b+m_s}\right) \\ 
&   \times \sin\theta^{12}_L \sin\theta^{23}_L  \sin\delta_L
\end{split}
   \\[3pt]
   \label{eq:Delta}
  \delta_R - \delta_L  &\simeq  s_a t_{2\beta} \frac{m_c+m_t \sin^2\theta_L^{23}}{m_s+m_d}
  \end{align}
  In the above formulas we keep small masses of the first generation in the denominators  in order to help the reader see the origin of such terms. It should be noted that the phase difference $\delta_R - \delta_L$ is accompanied with the $\sin\theta^{13}_L$ mixing angle. It suffices to change the signs of quark masses accordingly to get all the other solutions. The absolute values of the mixing angle differences are quite stable under these transformations, while the phase difference varies somewhat.
    
\begin{figure} 
  \hspace{-0.3cm}  \includegraphics[scale=0.515]{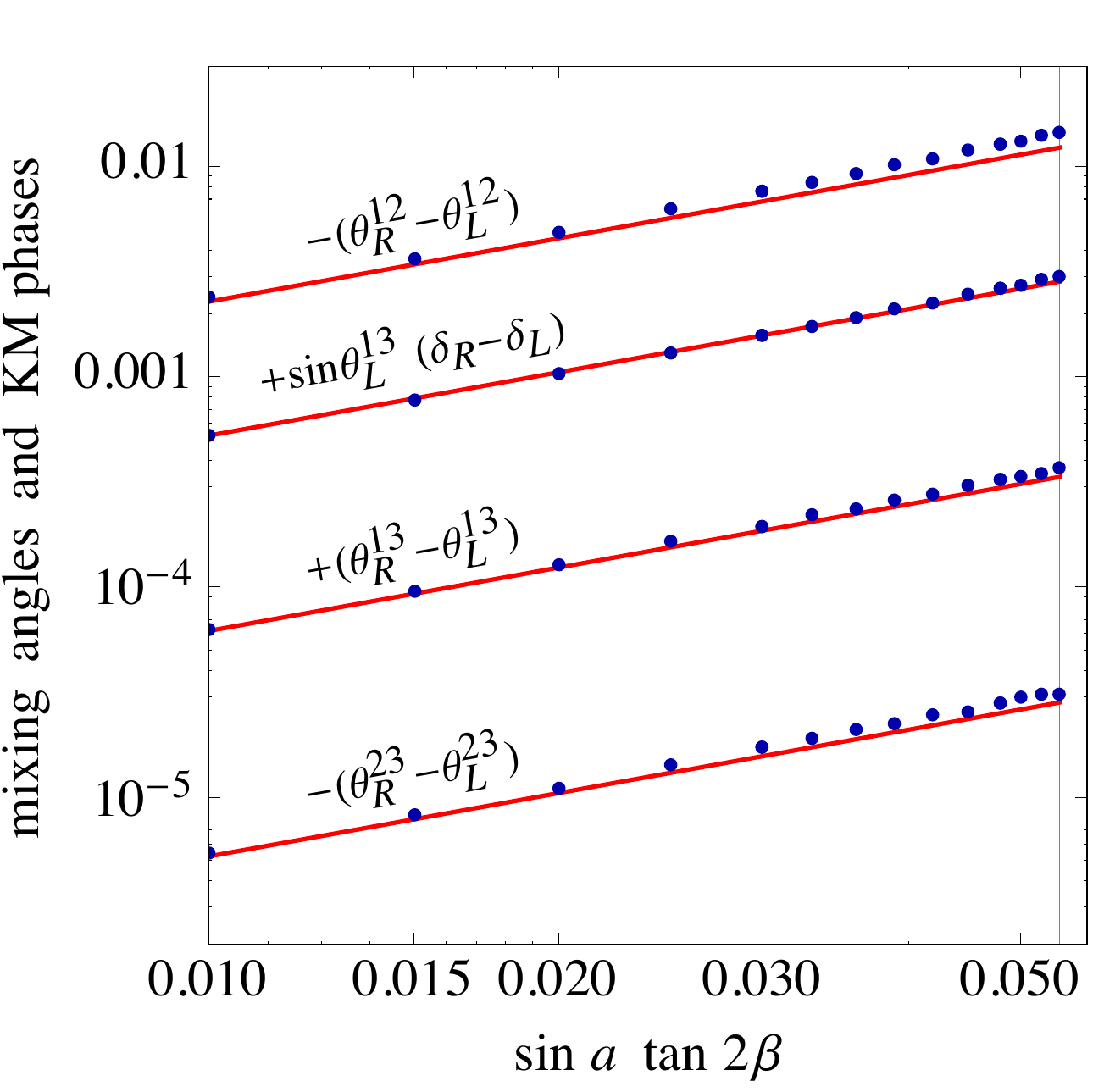} 
\caption{The differences between the right and left handed mixing angles and the KM phases ($\delta_R-\delta_L$ is multiplied with the accompanying $\sin \theta^{13}_L$). The first order terms are given by red lines, the blue dots denote exact numerical solutions. The agreement is striking in all of the allowed region $s_a t_{2\beta} \lesssim 0.055$. }
\label{fig:VR-angles}
\end{figure}

  Notice that the angle differences vanish in the limit of zero CKM phase $\delta_L$, the reason being that in this limit, the first order terms in $s_a t_{2\beta}$ in \eqref{eq:master} are purely imaginary and thus affect only the phases. It is clear that the angle differences are extremely small, suppressed by small mixings. In Fig.~\ref{fig:VR-angles}  we plot in red lines these first order results, and with blue dots the exact numerical solutions. The agreement between them is impressive; the first order is an excellent approximation. The difference between the 1-2 mixing angles is always less than about $10 \%$.

It may not be obvious why the differences of angles are always accompanied by other small mixings. The simple understanding of this important result comes, strangely enough, from the discussion of the non-realistic two generation situation.
Notice a surprising fact: at the first order the left and right mixing angles are equal, since there is no CP phase in $V_L$ (Cabibbo rotation is real) and the only change is in the imaginary components, i.e. phases. If this was to be true to all orders, it would be a rather useful result, for it would tell us that the difference between mixing angles in the three generation case must be proportional to the small CKM mixings, and thus is guaranteed to be small. Below we show that the equality of $\theta_L$ and $\theta_R$ is actually exact.
 
First, we compute the external phases from \eqref{eq:master}

 \begin{align}\label{eq:phases}
  \begin{split}
\omega_1&  \simeq  - \omega_3 +s_a t_{2\beta}   \frac{m_c}{m_s},
    \qquad \!\!\!\!
  \omega_2  \simeq - \omega_3 \simeq s_a t_{2\beta} \frac{m_t}{2m_b},
    \\[3pt]
     \omega_4&  \simeq  \omega_3- s_a t_{2\beta}   \frac{3 m_c}{2 m_s},
   \qquad\!\! 
     \omega_5  \simeq  \omega_3- s_a t_{2\beta} \frac{m_c}{2m_s}.
     \end{split}
      \end{align}

Unlike the expressions for the mixing angles and the KM phases, these phases depend strongly on the sign transformations that connect different solutions. The above formulas should just be taken as an example of all positive signs. There is one subtlety to keep in mind: in some cases the sign changes make the phases start from $\pi$ and not from zero, but that is easy to figure out.

\begin{figure}
\includegraphics[scale=0.5]{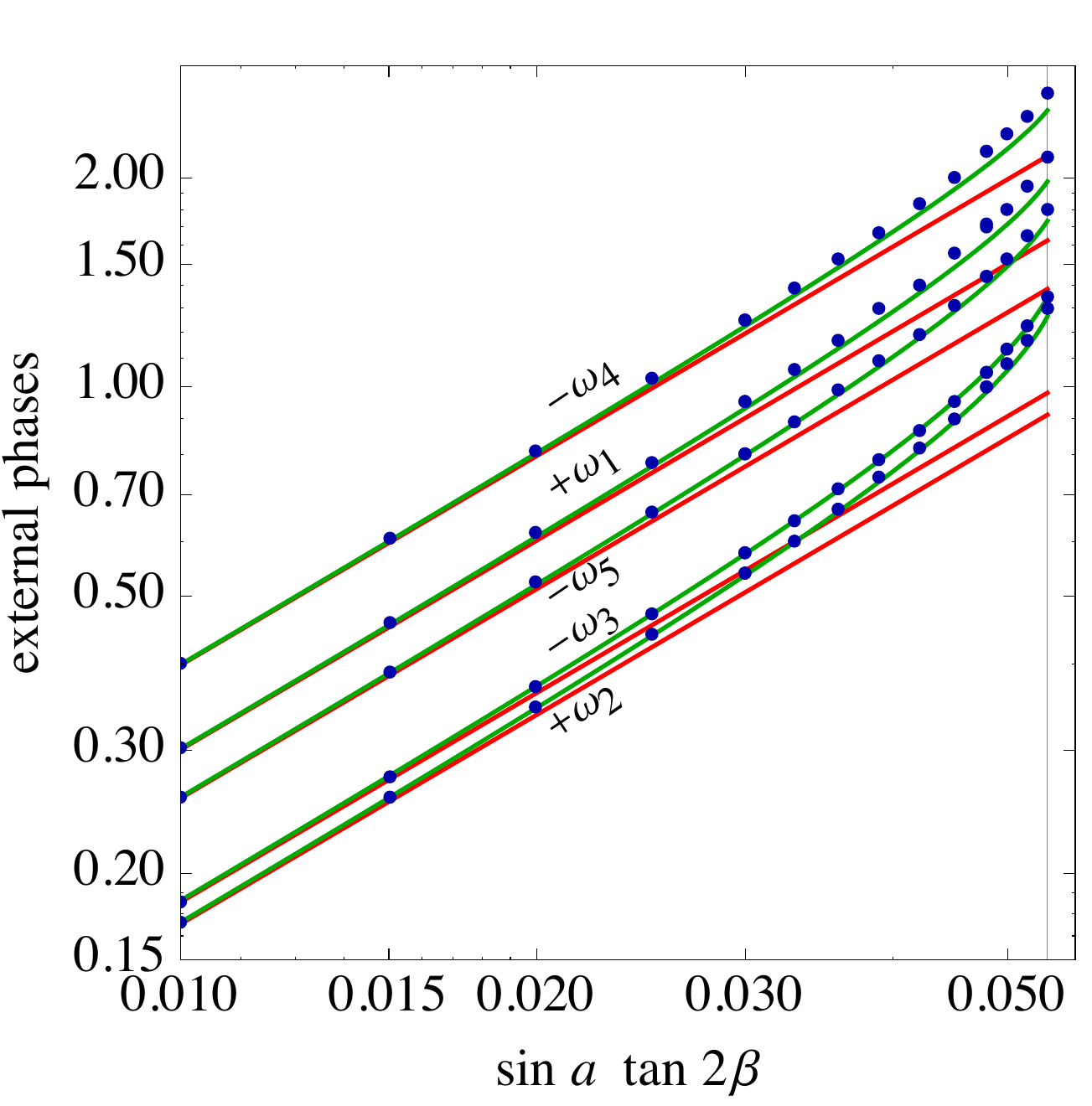}
\caption{Right handed external phases. Again, in red lines we give the first order expressions and with blue dots the exact numerical solutions; they  start diverging only for larger values $s_a t_{2\beta} \gtrsim 0.03 $. The inclusion of the higher order terms in green gives an excellent agreement with the exact results.
} 
\label{fig:VR-phases}
\end{figure}

 We plot these phases in Fig.~\ref{fig:VR-phases}. Again, the first order results are shown in red, and the exact numerical results in blue. Notice that in this case the results start diverging for larger values  $s_a t_{2\beta} \gtrsim 0.03 $, which simply means the lack of higher order terms in \eqref{eq:master}. It turns out that it is enough to use the following more precise form of $\omega_3$
 
 \begin{equation}\label{eq:omega3}
 \sin \omega_3 \simeq - s_a t_{2\beta} \frac{m_t}{2m_b}
 \end{equation}
 the rest remaining intact. The details of the computation are left for a longer paper to appear soon \cite{senjanovic:long}.
 In Fig.~\ref{fig:VR-phases} we also give in green the values for the above phases with $\omega_3$ given in \eqref{eq:omega3}; the agreement with the exact results is now excellent.


\subsec{RH charged gauge boson at colliders.} 
  The production strength of $W_R$ is proportional to
      $|(V_R)_{11}|^2$,  where from \eqref{eq:master} one has
\begin{align}\label{eq:VR11}
(V_R)_{11}\simeq  c_{12}\bigg(1-i s_at_{2\beta} \frac{m_c}{2m_d}s_{12}^2- \frac{s_a^2t^2_{2\beta}}{2} \frac{m_c^2}{4m_d^2} s_{12}^4 \bigg) 
\end{align}
   where $c_{12} \equiv \cos\theta_L^{12}, s_{12} \equiv \sin\theta_L^{12}$. 
 Since this is a high energy process, we used  \eqref{eq:master} at its face value, without any  scale dependence modifications.

      It is easy to see that  $|(V_R)_{11}|^2$ differs from $|(V_L)_{11}|^2$ by at most  a percent; thus to a great precision the production rate of $W_R$ is equal to that of $W_L$. Similarly, the decay rates into top and bottom, or di-jets  are in general to an excellent approximation given by practically the same left and right coupling. This is useful for direct searches for the LR symmetry at the LHC and illustrates clearly the importance of our result.

\subsec{  $K_L - K_S$ mass difference.}
This low-energy phenomenological example provides an indirect lower limit on the LR scale. The dominant effect comes from the charm quark in the box diagram with the $W_L$ and $W_R$ charged gauge bosons. Only the real part of the amplitude contributes and it is proportional to the following form of the mixing matrices 
\begin{align} \label{eq:KLKS-VRterm}
\begin{split}
&\hspace{-0.2cm}  \text{Re}[\mathcal{A}_{LR}(K\rightarrow \bar{K})]
  \! \propto \! \text{Re}\,[ (V_L)_{21}(V_L^*)_{22}(V_R)_{21}(V_R^*)_{22}]\\
&\hspace{-0.00cm}\simeq  c_{12}^2s_{12}^2\bigg[1-\frac{1}{2} \big(s_at_{2\beta} \big)^2 \bigg(\frac{m_c}{2m_s}  c_{12}^2+\frac{m_c }{2m_d}s_{12}^2\bigg)^{\!\!2} \bigg] 
\end{split}
\end{align}
This once again illustrates the usefulness of our findings. Previously, it was simply impossible to provide such analytic expressions. 

Naturally, there is a scale dependence when applying \eqref{eq:master} to low energies. For $W_R$ to contribute substantially to this process its mass should lie below 10 TeV or so, and  the scale dependence would amount a tiny effect for the case of the first two generations since their Yukawa couplings are small and there is no running due to QCD.  

 There is a plethora of both low and high energy processes with similar dependence on the elements of $V_R$ and their observation would pave the way for their eventual reconstruction \cite{Fowlie:2014mza}. It was tough enough to determine $V_L$ in the SM so this task is obviously a tremendous challenge, the discussion of which is beyond the scope of this work.

\subsec{Two generation case: exact RH angle.}
The $2\times2$ matrices possess a special property: the off-diagonal elements of the square root of a matrix are proportional to the off-diagonal elements of the matrix itself, with the same coefficient of proportionality. This simplifies matters greatly, and 
from the unitarity of $U_u$ (or equivalently  $U_d$) one immediately gets
  \begin{align}
  &\left|  (V_Lm_dV_R^{\dagger} ) _{12} \right|=\left| (V_Lm_dV_R^{\dagger} ) _{21} \right|
  \end{align}
 with  the unique solution (up to a sign ambiguity discussed below \eqref{eq:master}) of the same mixing angles
  \begin{align}\label{eq:left=right}
  & \theta_R= \theta_L
  \end{align}
This result allows for the understanding of the approximate equality between $V_R$ and $V_L$. As discussed above, the small parameter $s_a t_{2\beta}$ is sometimes accompanied by large quark mass ratios  and the product can be close to one, complicating matters. However, the difference between mixing angles must be proportional to other - fortunately small - mixing angles. Thus, the situation encountered in \eqref{eq:difference12}-\eqref{eq:difference13} continues to be true to all orders in $s_at_{2\beta}$.
    The near equality of left and right mixings is guaranteed by the smallness of left-handed quark mixing angles.

\Sec{IV. Summary and outlook. }

  In this Letter we have been able to elucidate the long-awaited form of the RH quark mixing matrix $V_R$ in the minimal left-right symmetric model
augmented with
  generalized parity. We found exact equations valid in all of the parameter space that allow for the numerical determination of $V_R$. We give the series expansion in terms of a small parameter that measures the departure from the hermiticity of the quark mass matrices. Moreover, we give a simple demonstration of the approximate equality between left and right mixing angles, using the important fact that in the two generation case $\theta_R=\theta_L$. The small CKM mixing angles then guarantee practically equal mixing angles in the realistic three generation case.  
  
 The case of $\mathcal{C}$ was easy to understand. Since quark mass matrices are symmetric, the mixing angles are equal, the KM phases have opposite signs while $V_R$ contains five extra arbitrary external phases. The situation with $\mathcal{P}$ is even more appealing: the mixing angles differ only slightly, while the phases in $V_R$ are calculable as functions of quark masses and a spontaneously induced phase.
         
It is a remarkable fact: the world in which we live ensures that the symmetry between left and right mixing angles remains practically exact, in spite of parity being broken almost maximally. 

 The knowledge of $V_R$ is crucial for the production and decays of the RH charged gauge boson, the decays of the heavy scalar doublet in the bi-doublet, the prediction of the neutrinoless double beta decay, the calculation of the strong CP parameter and $K$ and $B$-meson physics, to name a few. We gave here an example of  the $W_R$ production and the  $K_L - K_S$ mass difference for the sake of illustration and we will discuss other processes at length in a longer paper now in preparation.

   \vskip 0.3 cm

   We are grateful to Alejandra Melfo and Miha Nemev\v sek  for their collaboration in the initial stages of this work and to Darius Faroughy for checking most of our results. We wish to thank Federica Agostini, Alejandra Melfo, Fabrizio Nesti, Natasha Senjanovi\'c and Yue Zhang for numerous discussions and comments, and for careful reading of the manuscript.


\end{document}